\documentstyle[12pt]{article}
\newcommand{\bib}{\bibitem}

\newcommand{\spz}{\hspace{0.7cm}}

\newcommand{\nn}{\nonumber}
\newcommand{\fr}{\rightarrow}
\newcommand{\de}{\partial}
\newcommand{\ri}{\right}
\newcommand{\lf}{\left}
\newcommand{\w}{\wedge}

\newcommand{\ga}{\gamma}
\newcommand{\Ga}{\Gamma}

\newcommand{\eq}{\begin{equation}}
\newcommand{\en}{\end{equation}}
\newcommand{\bea}{\begin{eqnarray}}
\newcommand{\eea}{\end{eqnarray}}

\newcommand{\ba}{\begin{array}}
\newcommand{\ea}{\end{array}}

\newcommand{\virg}{\spz,}
\newcommand{\pu}{\spz.}

\newcommand{\resection}[1]{\setcounter{equation}{0}\section{#1}}

\newcommand{\hg}{\hat{g}}
\newcommand{\hl}{\hat{l}}
\newcommand{\half}{\frac{1}{2}}
\newcommand{\ttrd}{\frac{2}{3}}

\newcommand{\M}{{\hbox{\large $\cal M \;$}}}

\newcommand{\T}{{\hbox{$Tr\;$}}}
\newcommand{\G}{{\hbox{\large $\cal G \;$}}}

\newcommand{\cH}{{\hbox{\large $\cal H \;$}}}
\newcommand{\AP}[1]{Ann.\ Phys.\ {\bf #1}}
\newcommand{\CMP}[1]{Comm.\ Math.\ Phys.\ {\bf#1}}
\newcommand{\FAP}[1]{Funk.  \ Anal. \  Priloz. \ {\bf#1}}
\newcommand{\LMP}[1]{Lett. \  Math. \  Phys. \ {\bf#1}}
\newcommand{\IJMP}[1]{Int.\ J.\ Mod.\ Phys.\ {\bf#1}}

\newcommand{\JMP}[1]{J.\ Math.\ Phys.\ {\bf#1}}
\newcommand{\MPL}[1]{Mod.\ Phys.\ Lett.\ {\bf #1}}
\newcommand{\NC}[1]{Nuovo Cim.\ {\bf #1}}
\newcommand{\NP}[1]{Nucl.\ Phys.\ {\bf #1}}
\newcommand{\PL}[1]{Phys.\ Lett.\ {\bf #1}}
\newcommand{\PR}[1]{Phys.\ Rev.\ {\bf #1}}
\newcommand{\PRL}[1]{Phys.\ Rev.\ Lett.\ {\bf #1}}
\newcommand{\TMP}[1]{Teor.\ Math.\ Phys.\ {\bf #1}}

\begin{document}
%%%%%%%%%%%%%%%%%%%%%%%%%%%%%%%%%%%%%%%%%%%%%
%%%%%%%%%%%%%%%%%%%%%%%%%%%%%%%%%%%%%%%%%%%%%
\setlength{\unitlength}{.8mm}

\hfill  DFUL-1/06/95\\ \vskip 1cm
 \begin{center}  {\large \bf
 TOPOLOGICAL  FIELD  THEORY} \\[.5cm]
{ \large \bf AND
NONLINEAR $\sigma$-MODELS }\\[.5cm] { \large \bf
ON
 SYMMETRIC   SPACES }\\[.5cm]  \vskip 1.2cm
 {\large  \it
L. Martina \footnote{E-mail: martina@le.infn.it}
	     , O. K. Pashaev$^{\dagger \,*}$ \footnote{E-mail:
pashaev@main1.jinr.dubna.su} and G. Soliani \footnote{E-mail:
soliani@le.infn.it}}\\

\vskip .8cm
  {\it Dipartimento di Fisica dell'Universit\`a and INFN
Sezione di
 Lecce}\\
 {\it 73100 Lecce, Italy}\\ $\dagger)$ {\it Joint Institute for
Nuclear Research, 141980 Dubna,
 Russia}  \footnote{ Permanent address}\\ $*)$ {\it Department of
Mathematics, TUBITAK-Marmara
 Research Center}\\
 {\it 41470 Gebze-Kocaeli, Turkey}  \\ \vskip 1cm
\noindent {\bf Abstract}
\end{center}

{\it  We show that  the classical non-abelian pure
Chern-Simons action is  related  to
nonrelativistic  models in (2+1)-dimensions, via
reductions of the gauge connection in  Hermitian
symmetric spaces. In  such models the matter fields are coupled to
gauge Chern-Simons fields, which are associated with  the isotropy
subgroup of the considered symmetric space. Moreover, they can be
related to  certain (integrable and non-integrable)  evolution
systems,  as  the Ishimori and the Heisenberg model.  The  main
  classical and quantum properties of these systems are
discussed  in connection with the topological field theory and the
condensed matter physics.}
\vskip 1cm

\newpage  \section{Introduction}  In the last years,
a new class of general covariant field theories,
called  quantum topological field theories (TFT's),
were introduced ~\cite{w1,w2}.These models  were
originally  related to  Yang-Mills instantons
{}~\cite{w1}, $\sigma$-models~\cite{w3} and
gravity~\cite{w4}.  Successively,  the work by  Witten
established a relationship between the
 covariant four-dimensional quantum fields and
the  abelian and non-abelian  quantum
Chern-Simons (CS) action on a three-dimensional
manifold \M.  The most relevant property of these
theories is that of having  observables (the Wilson
line operators)  which are metric independent.
Moreover, in this framework, the vacuum
expectation values of such  observables are
invariant under smooth deformations on \M,
i.e.  they are topological
 invariants of closed links in \M and are   related to
the Jones  polynomials in knot theory~\cite{jo}  and
their generalizations. \par From another  point of  view,
models of point particles coupled to a CS  gauge field
in 2+1 dimensions, have gained attention owing to
their  peculiar long range interaction ~\cite{Ha}. In
particular, in the abelian CS theories the
interacting point-particles  ( anyons) obey  a
fractional statistics  ~\cite{Le,Wil}  and play a role
in  fractional quantum Hall effect and in high
temperature superconductivity ~\cite{Pra,Lau}.\par Such a  kind
of theories can be extended to the non-abelian case ~\cite{bak}
and  can be treated as non-relativistic quantum systems
{}~\cite{jak,japi1,japi2,japi3} . The quantization
procedure leads to an N-body Schr\"odinger equation,
with the Aharonov-Bohm potential in the abelian
case, and with the presence of   the
Knizhnik-Zamolodchikov connection in the non-abelian
case ~\cite{bak}. Moreover, in both   cases, within the
classical field approach  equations of the gauged
nonlinear Schr\"odinger (NLSE) type arise. In the  static
self-dual situation, such systems become the Liouville
equation and its  integrable multicomponent
generalizations  ~\cite{jak,japi1,japi2,japi3,jakpiSupp}.
 These results  lead
directly to a link  between the static CS-theories
 and the static reductions of
integrable equations in (2+1)- dimensions,  like the Ishimori
model and the Davey-Stewartson equation \cite{self}.

Now, the basic
property of the integrable equations is that they can  be represented
as zero-curvature conditions for certain special linear connections
(the Lax pair operators) in suitable spaces. Incidentally, this is a
representation shared by the CS equations of motion in a
three-manifold. Furthermore,  among the great  amount of
consequences which derives from this fact, a particular role is
plaied by  the transformation properties between two of such
integrable equations. For example, it is well known that the NLSE is
"gauge equivalent" to the Heisenberg  model  in 1+1 dimensions
{}~\cite{ZakTa,Ku,MMP}. Furthermore, the method of the gauge equivalence
has been  extended to integrable (2+1)-dimensional equations, for
instance between the   Ishimori   and the Davey-Stewartson equation
(integrable extensions in (2+1)-dimensions of the Heisenberg model and
of the NLSE, respectively)   ~\cite{Lip,kon,noi,mik}. However, this
method can be made suitable  for handling non-integrable systems
also,   at least under certain conditions. More specifically, the
method is based on the interpretation of the spin  variables as
elements of a coset space {\G/\cH}. Then,  the geometric
characterization  of such a  space   is given in terms of the
zero-curvature condition for  the  associated chiral currents.

In
many  physical applications ~\cite{pasmak}, the  coset space
{\G/\cH} is taken to be a symmetric space,  with  isotropy group {\cH}
 \cite{Dub,Hel}.  Hence,  the current components,  taking
values in the Lie algebra of  \cH,  are considered as  the
local  gauge  fields of the theory.  The  further
degrees  of freedom belongs to the tangent space of the symmetric
space and are interpreted as the "matter fields".   When this
procedure is applied to a specific spin model, we refer to it as the
"tangent  space representation" of the given theory.  In  this
formalism   the spin dynamics provides further restrictions  on  the
chiral currents,    leading  to  some multicomponent
gauged NLSE's.   The integrable systems
are recoverd  only for a particular choice of the gauge
connection.\par In particular,  by using this scheme  we
have  mapped the   continuous
Heisenberg  model in (2+1)-dimensions   into an abelian
pair of  CS gauged  NLSEs \cite{CSF}. Thus, we have
proved the existence of a connection between
continuous ferromagnetic systems and CS models.

At this point, we notice that the classical field equations
for the pure CS theory are formally just the
zero-curvature conditions found in our tangent space
approach to the planar ferromagnets. The only
difference concerns the local gauge group, which  is  the whole  \G
for the pure case,  and the subgroup \cH  for ferromagnets. Therefore,
in the last case one has  supplementary constraints, which are
\cH-invariant by construction, and  break the general
\G-invariance.

These considerations have suggested us to study the CS
theory in the symmetric space \G/\cH.  Following this idea,
in Section II  we  study such type of reductions and explain in which
sense we obtain certain  matter fields coupled to a local gauge  CS field.
Therefore, we show  how  one  can  introduce  suitable classical
"gauge-fixing" conditions, which have to be  invariant under gauge
transformations on \cH.
Furthermore, the
existence of a  classification for  the symmetric spaces
\cite{Dub,Hel},  and  the requirement  that  the gauge fixing
condition has to be  invariant under specific subgroups,
implies  a   classification of the possible theories  obtained by
using the outlined procedure.  In Section III  we study the generalized
$\sigma$-models in the tangent space formalism. Thus, we  show
the deep relationship which  exists  between the spin models
and the topological field theories. This relationship
is   studied in detail in Section IV,  where the  classical canonical
structure of the
$SU\lf(2\ri)/U\lf(1\ri)$ model is given. We show  also how
the constraints of the theory can be solved,  and how  the choice of
the Heisenberg model as  gauge-fixing condition yields a hamiltonian
system,  in which the $U\lf( 1 \ri)$
gauge-invariant  and  the  non-invariant
variables are separated. In Section V we discuss the quantization of
the previous model and its connection with the multianyonic systems.
Finally, Section VI  is devoted to some remarks and   open problems.
    \section{Chern-Simons theory on symmetric
spaces}
 The Chern-Simons
action for a compact non-abelian   simple Lie group \G   on   an
oriented  closed three-dimensional  manifold \M
is
 \eq
 S[J] = {k \over {4\pi}}\int_{\M} \T (J\wedge dJ +
\frac{2}{3} J\wedge J\wedge J)
 \label{1}
\en
where  $J$ is  the  1-form
 gauge connection with values in the  Lie  algebra
$\hat{g}$ of \G and the trace is taken in a chosen
representation.   The action  (\ref{1})  is
 manifestly invariant under general coordinate
transformations (preserving orientation and
volumes). Moreover,
 under a  generic gauge map $G: \M  \fr \G$ the gauge
connection transforms as usual by
 $
 J \fr G^{-1}JG + G^{-1} dG
$ .
Correspondingly,  the  action (\ref{1})  changes as
$
 S[J] \fr S[J] +  2 \pi\, k\, W(G) ,
 $
 where
 \eq
 W(J) = {1 \over 24\pi
^{2}}\int_{\M}\T (G^{-1}dG\wedge G^{-1}dG\wedge
G^{-1}dG)
 \en
 is the winding number of the map $G$ and takes
integer values, because of  the
result  $\pi _{3}(\G) = {\bf Z}$  of  the  homotopy theory
\cite{Dub} .  The gauge  invariance  of the  quantum
theory  (defined by  the functional integral
approach \cite{w1}-\cite{w4},  or in the canonical
formalism \cite{dun}) implies  the quantization
 of   the constant $ k$. Nevertheless, the  infinitesimal
gauge transformation
 $G = e ^{ \lambda}  \simeq  I +
 \lambda $ ($\lambda: \M  \fr {\hat{g}}$ ), acting on  $J$ as
 \eq
 \delta J = [J,\lambda ] + d\lambda  \virg
\label{gauge1}
 \en
 leaves the action  invariant.
 The classical equations of motion for the action (\ref{1})
is the  zero-curvature condition  \eq
 F = dJ + J\wedge J = 0 \pu \label{eqmot}
 \en

Now,  let us  choose in \G a closed subgroup \cH, such that the  Lie
 algebra $\hg$ of \G  satisfies the  so-called ${\bf Z}_{2}$ - graduation
condition
 \eq
\hg = \hl^{(0)} \oplus
\hl^{(1)}  \virg \qquad
 [ \hl^{(i)}, \hl^{(j)}] \subset  \hl^{(i+j)\; {\hbox{mod (2)}}  } \virg
\label{z2g}
\en
 where
$\hl^{(0)}$  is the Lie algebra of
 \cH and $\hl^{(1)}$  is the  (vector space)
complement of  $\hl^{(0)}$ in $\hg$  \cite{Dub,Hel}. The group \G acts
transitively on the coset space \G/\cH, which   can be identified with a
differential manifold belonging to the class of the symmetric spaces.
The subgroup \cH  leaving invariant the  points of  \G/\cH  is called the
isotropy group.    At any  point
$p_{0}   \in$
\G/\cH,  the  tangent
 space $T_{p_{0}}\left({\G/\cH }\ri)$
is isomorphic to   $\hl^{(1)}$ and a  torsion-free connection can
be  canonically defined. Moreover,   we can induce in  a  natural way a
riemannian metric on
$T\lf({\G/\cH }\ri)$, by restricting
the Killing   form on $\hg$  to $\hl^{(1)}$.   However,  for our aims  we
are  interested in those symmetric spaces which  enjoy    a complex
structure, the so-called   Hermitian symmetric spaces
\cite{Hel},  the main  algebraic properties of which  are:
\par \noindent 1)  there exists an element  $   A $  belonging to
the   Cartan subalgebra of  $\hg$,  such that  its
centralizer in $\hg$ coincides with  $\hl^{(0)}$, thus
$\left[ A, \hl^{(0)} \right ] = 0$  holds as  a  particular case;\par
\noindent  2)  one can write  $\hl^{(1)} =  \hl^{(1+)} \oplus  \hl^{(1-)}$,
where
$\hl^{(1 \pm )} =  \hbox{\rm span}_{\alpha \in \Phi^{+}}  \left\{{ e_{\pm
\alpha} }\right\}$  are expressed by using  those  elements   of the
the Weyl-Cartan
basis, which correspond to a subset $\Phi^{+}$ of the positive root
system   on which  $\alpha  \lf( A \ri)$ is a  constant, thus    $\left[ A,
\hl^{(1
\pm)}\right ] =  \pm \alpha  \lf( A \ri)
\hl^{(1 \pm)}$ holds; \par \noindent 3)
by using a proper scaling,
$ad\, A$ can
be considered as a linear involutive endomorphism  on   $\hl^{(1)}$
supplying the complex structure on it ;\par \noindent
4) the commutation relations
$\left[{{  e}_{\pm \alpha },{e}_ {\pm \beta }}\right]=0$  hold for all
pairs of basis elements in $\hl^{(1 \pm )}$.

\par  By using the   graduation  (\ref{z2g}),  we can first   assume
that the current $J$  has  the form
\eq
J = J^{(0)}+ J^{(1)} \virg      \label{dec}
\en
where $J^{(i)} \;$  are 1-forms  taking  values in
$\hl^{(i)}$. Hence, by resorting to   the  properties 1) -
4),   the  CS  action (\ref{1})   becomes   \eq
 S[J^{(0)}, J^{(1)}] = {k \over 4\pi }\int_{\M} Tr(J^{(0)}\wedge dJ^{(0)}
+
\ttrd J^{(0)}\wedge J^{(0)}\wedge J^{(0)}
+ J^{(1)}\wedge \hat{\rm D}
J^{(1)})
\label{2}
 \en
 where $\hat{\rm D} = d + \cdot
\wedge J^{(0)}+ J^{(0)}\wedge  \cdot \,$ is the
covariant exterior
 derivative.  The expression (\ref{2}) suggests  us to
interprete
 $J^{(0)}$  as a CS-gauge field and $J^{(1)}$ as a
coupled matter field.   In this
 sense  we  have reformulated a   \G-invariant
 pure non-abelian CS theory (\ref{1})
  as   an interacting matter gauge field
 theory   with  group \cH. The whole set  and the  properties
of such theories are directly connected with the
classification problem  of all possible Hermitian  symmetric spaces,
which is  completely solved \cite{Hel}.   Moreover,
although at   first glance  the   distinction between
matter and gauge fields could seem rather  artificial, it is
preserved  under the action of \cH. In fact, keeping in mind the
infinitesimal transformation (\ref{gauge1}) and  the ${\bf
Z}_{2}$-graduation, with   $\lambda =
\lambda^{(0)} + \lambda^{(1)}$, we obtain
 \bea
 \delta J^{(0)} = [J^{(0)},\lambda ^{(0)}] + d\lambda
^{(0)}  \nn \\
 \delta J^{(1)} = [J^{(1)},\lambda ^{(0)}] \pu \label{gH}
 \eea
for $\lambda^{(1)} =0$.
Anyway, the action (\ref{2}) still possesses the property of being
\G-invariant. Then, we  have the freedom to introduce certain gauge
fixing conditions, which are invariant under   \cH.  This procedure could
yield specific field   models.

 In order to display  some concrete examples of such models,  it is
convenient to follow a suggestion by
 Witten
\cite{w2}. This consists  in  "chopping"  a general 3-manifold \M   into
pieces, each  of  them   is  isomorphic
 to  $\Sigma  \times {\bf R}$, where $\Sigma$ is a
Riemann surface  and     $ {\bf R} $   is interpreted
 as the time.
 In doing so, the
exterior derivative  $d$ and the
current $J$   are    expressed  in terms of   time  and
space
 components
 \eq d = d_{0}\;+ {\bf d} \qquad \lf( d_{0} = dx^{0}\,{\de}_{0}\ri) \en
and
\eq
    J =  {\it A}_{0}+ {\bf A} =  {\it A}_{0}^{(0)} +
{\it A}_{0}^{(1)} + {\bf A}^{(0)} + {\bf A}^{(1)} \virg \label{comp}
\en
 respectively.  In the last equation   we have taken into
account the ${\bf Z}_{2}$-graduation (\ref{dec}) for $J$,   ${\bf
A}^{(i)} = A_{a}^{(i)}\,dx^{a}$ is a real connection on $\Sigma$ and
${\it A}_{0}^{(i)} =  A_{0}^{(i)}\,dx^{0}$ are 1-forms on ${\bf R}$ .
Furthermore,  at this point  we   parametrize  $\Sigma$ by means
of the local complex  coordinates
 $
 z = x_{1}+  i\, x_{2} , \;\;\bar{z} =  x_{1} -  i\, x_{2}
 $ .
Now,  we can consider  tensors, either of
covariant or contravariant type,  with some
definite number of holomorphic and anti-holomorphic indeces.  For
instance, the  cotangent space of  $\Sigma$  will
be decomposed into the direct sum of two subspaces,  one of
them is spanned by
$dz$,  and   the other by $d\bar{z}$. An important observation is
that barred and unbarred tensors do not transform into each
other  under a holomorphic change of coordinates. Specifically, our
connection $J$ can be rewritten  in the form
\eq
J =  V_{0} +  V + \bar{V}  + M_{0} + M +  \bar{M} \virg
\label{VMdecomp} \en
where
\bea
V &= \half  (A_{1}^{(0)}- i A_{2}^{(0)})\; dz \;,\qquad  &
 \bar{V}  =  \half  (A_{1}^{(0)} + i A_{2}^{(0 )}) \; d\bar{z}   \virg \\
M &= \half  (A_{1}^{(1)}- i A_{2}^{(1)})\; dz \;, \qquad &
 \bar{M}  =  \half  (A_{1}^{(1)} + i A_{2}^{(1) }) \; d\bar{z}   \virg
\eea
and  we    put
\eq
V_{0} = {\it A}_{0}^{(0)}  \virg   M_{0} = {\it A}_{0}^{(1)}
\en for brevity.
Correspondingly,  we can  split the exterior derivative  in the form
\eq
 {\bf d} =  \de  + \bar{\de } =
dz\;   {\de}_{ z} + d\bar{z} \;  {\de }_{ \bar{z}}  \virg
\en
where $\de$ e $\bar{\de}$  are
  holomorphic and
anti-holomorphic operators  globally  defined on $\Sigma$.

Taking  into account this   construction on
 $\Sigma  \times {\bf R}$,  the
covariant exterior derivative introduced in the action  (\ref{2}) is
written  now as  the sum of three terms: $ D =  {\cal D} +   \bar{{\cal
D}} +
  {\cal D}_{0} $, where
\bea
{\cal D} =& \de  + { V}\wedge
 \cdot  + \cdot \wedge { V} ,\qquad
 \bar{{\cal D}} = \bar{\de } + \bar{{ V}}\wedge  \cdot +
\cdot \w \bar{{ V}},\nn \\
  &{\cal D}_{0} = d_{0}   + {
V}_{0}\w  \cdot + \cdot \w { V}_{0} .   \eea

Then,  just  by rearranging   the  $\cH$-
invariant  action (\ref{2}) in the complex variables, we obtain   the
canonical Darboux form
 \bea
 S =  \frac{k}{4\pi}   \int_{\Sigma  \times {\bf R}}
&\T ( { V}\w  d_{0}\, \bar{{ V}} + \bar{{ V}}\w
d_{0}\,{ V} + { M}\w d_{0}\,\bar{{ M}} + \bar{{
M}}\w d_{0}\, { M} + \nn \\
  & + 2 { V}_{0}\w \lf({\de}\bar{{ V}} + \bar{{\de}}{
V} +  { V}\w  \bar{{ V}} +  \bar{{ V}}\w { V}  +
M \w \bar{{ M}} + \bar{{ M}} \w M \ri)
\nn \\ & + 2 { M}_{0}\w ({\cal D}\bar{{ M}} + \bar{{\cal
D}}{ M})) \pu
                      \label{3}\eea

The corresponding equations  of motion are

\bea {\de  } \bar{V}+\bar{{\de  }}
V + V\w \bar{V} + \bar{V}\w V &=& - \lf( M\w
\bar{M} + \bar{M}\w M \ri),  \label{eqm1}\\
 d_{ 0}V  +   {\de}   {V}_{0}  + V\w
{V}_{0} + {V}_{0}\w V &=& - \lf({M}_{0}\w M + M\w
{M}_{0}\ri) ,   \label{eqm2} \\ d_{
0}\bar{V} + \bar{   {\de}  }
{V}_{0} + \bar{V}\w {V}_{0} + {V}_{0}\w
\bar{V} &=& - \lf({M}_{0}\w
\bar{M} + \bar{M}\w {M}_{0}\ri) ,  \label{eqm3}\\
{\cal D}   \bar{M}+\bar{   {\cal D} }M& =&0 \virg  \label{eqm4}
 \\
 {   {\cal D}}_{0}M+   {\cal D}{   M}_{0 }&=& 0 \virg  \label{eqm5}
 \\
 {   {\cal D}}_{ 0}\bar{M}+\bar{{\cal
D}}{ M}_{0}&=& 0  \pu  \label{eqm6}  \eea

{}From  the expression (\ref{3})    the structure of  the Lagrange
density  with contraints is manifest. In particular,  $V_{0}$ and
$M_{0}$  are the Lagrange multipliers enforcing the
Gauss-Chern-Simons (GCS) law and  a sort of generalized self-dual
condition, which derives from the
torsion - free property of the
\G/\cH  manifold.  Furthermore,   by using the  infinitesimal
transformations (\ref{gH})
and the  expression (\ref{VMdecomp}), one  easily
sees  that  the action (\ref{3})
and the corresponding equations of motion (\ref{eqm1} - \ref{eqm6})  are
explicitly gauge invariant  under infinitesimal transformations
defined on \cH, in the sense that they  are not only a subclass of
the gauge symmetries admitted by the original model, but
preserve the decomposition
among    gauge fields  $V$-type  and   matter fields $M$-type.    On
the other hand, this picture
induces us to figure out  certain situations,  in which  the
invariance  under \G-transformations is  broken,  in such a way that
only the  \cH-invariance is preserved.  In other words,  we may
choose
 a  supplementary constraint  among the fields, which
 is   invariant under  the  \cH- transformations.    Since
such a constraint  have to transform accordingly  to
(\ref{gH}),    its general form  is
 \eq
\Ga \lf[M_{0}, M, \bar{M}, {\cal D}{   M}_{0 }, {\bar{{\cal D}}}{
M}_{0 }, \cdots  \ri] = 0   \virg \label{gf}
\en
where $\Ga$ denotes  an   arbitrary  differentiable  function,
depending on the $M$-type fields and on their covariant
derivatives.  If  Eq. (\ref{gf} ) can be explicitly solved for $M_{0}$,
then  we    replace  this  into the equations of motion
 (\ref{eqm1} - \ref{eqm6}),
obtaining nonlinear evolution equations for the matter fields $M$
and $\bar{M}$. The same substitution in the action (\ref{3}) leads to
a functional defined on the submanifold of \G / \cH  determined
by  Eq.  (\ref{eqm4}), which is seen as a   further constraint.
Conversely, in the case in which $M_{0}$ cannot be explicitly
determined, we can include the constraint (\ref{gf}) into the action
(\ref{3})  by means of a suitable Lagrange multiplier.    After all we
have  obtained     special  nonlinear
 evolution classical  models
for  the matter fields  in interaction with  the  CS
field (generally non-abelian).
Now,  carrying out   specific
calculations in TFT  one needs to break the
gauge symmetry , usually  by  the Weyl gauge
$A_{0} = 0$ \cite{w2}. Thus   the
idea we want to stress is that other suitable
constraints  of the form
  (\ref{gf}) can  be used, first because they may  help in    more
effective  calculations.  In particular,
when   Eq. (\ref{gf})   is  related
to certain integrable systems, for which exact solutions can be
given  analitycally  at least at
the classical level. Secondly, they can provide
integrable deformations of the topological symmetry of the original pure CS
model.
 Furthermore,   from Eqs. (\ref{eqm1} - \ref{eqm3})  we obtain the identity
\bea \left({\rm d{x}_{0}{\partial
}_{0}+\left[{{V}_{0},\cdot }\right]}\right)& \w  \left({M\w
\bar{M} +\bar{M} \w M}\right) + \left({{\partial
}_{}+\left[{{V}^{},\cdot }\right]}\right) \w \left({{M}_{0}\w
\bar{M}+\bar{M}\w {M}_{0}}\right)    \nn \\  &
+\left({{\bar{\partial }}_{} + \left[{{\bar{V}}^{},\cdot }\right]}\right)
\w \left({{M}_{0}\w M-M\w {M}_{0}}\right)=0,
\label{cont} \eea where we define   the "commutator"
operator  over 1-forms $\left[{A ,\cdot }\right] \w B = A \w B - B
\w A$ yielding a 2-form. Eq.  (\ref{cont}) yields  a continuity
equation, when a given  model is defined by the specific
dependence of $M_{0}$ on  $M$ and $\bar{M}$. The  conserved
density is given by the 2-form $\rho  dz  \w  d\bar{z} =
   {M\w
\bar{M}+\bar{M}\w M} $,  which furnishes  the set of
non-abelian conserved charges ${\cal Q} =  \int_{\Sigma}\,{\rho}\,
dz  \w  d\bar{z} $.

 A quite special situation occurs when we   consider the
Grassmann  manifold  $\G / \cH = SU(n+m)/S(U(n)
\times U(m))$  ({\it AIII} in the Helgason's  notation
\cite{Hel}).    In particular,    for  $m = 1$  we obtain    the
complex projective   $CP^{  n }  = SU\lf( n + 1 \ri)/S \lf( U \lf(
n \ri)   \times U \lf( 1\ri) \ri)  $ model.  In general,    the
gauge connection
$J$ is anti-hermitian  ($ J^{\dagger} = - J$ )
 and decomposes in the block  form
\eq
J=\left({\matrix{i {V}^{\left({n}\right)}& R\cr
Q &i {V}^{\left({m}\right)}\cr}}\right) \virg \label{aiii}
\en
 where  $ V^{(l)} = v_{\mu}^{(l)} dx^{\mu}$
are
 $l \times   l $ matrix valued one-forms   with $l =n, m$, $Q =
q_{\mu} dx^{\mu}$ and $R =  r_{\mu} dx^{\mu}$ are $ m \times  n$
and $n \times  m$
 matrix valued one-forms,  respectively.  The anti-hermiticity
of $J$ implies that   $v^{(l)  \, \dagger}_{\mu}
= v^{(l)}_{\mu}\; \lf( l = m, n
\ri) $ and $ r_{\mu} = -  q^{\dagger}_{\mu}$ .  Thus,   in   the complex
formulation we  introduce the forms
  \bea \matrix{{\psi }_{-} dz=&{1 \over
2}\left({{q}_{1}-{i q}_{2}}\right)dz \qquad &, &{\psi }_{+}
d\bar{z} =&{1 \over 2}\left({{q}_{1}+{i
q}_{2}}\right)d\bar{z}\virg\cr {v}^{\left({n}\right)} dz = &{1 \over
2}\left({{{v}^{\left({n}\right)}}_{1}-{i{v}^{\left({n}\right)}}_{2}}\right)
dz &,&{v}^{\left({m}\right)} dz=&{1 \over
2}\left({{{v}^{\left({m}\right)}}_{1}-{i {v
}^{\left({m}\right)}}_{2}}\right) dz ,
\cr} \label{cp}    \eea  which  give
a block decomposition of  the matter   and
the gauge    fields, $M$ and $V$, respectively. Then, substitution into
(\ref{3}) yields  the action  in
terms of $\psi_{\pm}$ and $v^{(i)}$.   We shall
  not   write  this    for brevity.   Here we  remark only that  in the
$CP^{ n}$  case  the matter fields
$\psi_{\pm}$ are $n$-component row vectors. Moreover,   the
cubic  nonlinear  self-interaction for the abelian $U(1)$  field
 vanishes  identically in the action. This fact  implies that the
corresponding  magnetic field $ B^{\lf(1\ri)} = \epsilon^{ij}\,
\de_{i}\,v_{j}^{\lf(1\ri)}$ is proportional to the charge density
$\rho^{\lf(1\ri)} = -  i \lf(\psi_{+}    \psi_{+}^{\dagger} - \psi_{-}
\psi_{-}^{\dagger}\ri) $.  Then the particles with electric charge
${\cal  {Q}}_{el} = \int_{\Sigma}\, \rho\,  dz \w d\bar{z}$ are also
flux-tubes with total magnetic  flux $\Phi_{m} = - {\cal {Q }}_{el}$ .
  But  for  $n \ge 2$,  accordingly to  the equations of motion
(\ref{eqm1}), the "coloured" magnetic field  ${B}^{(n)} = \lf[
\partial_{z}
 {\bar{v}}^{(n)}- \de_{\bar{z}}{v}^{(n)}
 +i\left({v}^{(n)}
{\bar{v}}^{(n)} - {\bar{v}}^{(n)}
{v}^{(n)}\right)
\ri]  = -  i \lf(\psi_{+}^{\dagger}   \psi_{+} - \psi_{-}^{\dagger}
\psi_{-}\ri)    $
 appears. However,  electric and "coloured" charges are not
completely independent, since in general   contributions  from the
non-abelian fields
to the spatial components of the electric current  are present  (see
Eq. (\ref{cont})).  Nevertheless, it is remarkable that the models of
non-abelian CS   field  coupled to the matter
\cite{bak} can be embedded into the scheme of our
$CP^{n}$ models, giving a specific example of constraint (\ref{gf}).

\resection{   Generalized $\sigma$-Models in the tangent space}

In this Section we show how
specific models  can be embedded into the  theory  developed  in
Section II.     In particular, we deal with
  generalized classical spin models,  whose spin phase
space is a  symmetric space. There are  several  motivations  for
studying  these type of systems. Many of them   can be found  in
\cite{pasmak} and    \cite{Salam}.

First of all  we
restrict ourselves   to the 2 + 1 dimensional
generalized Heisenberg model, which is a  natural
extension  of 1+1 dimensional integrable version
defined on   symmetric spaces.  As  is well known from a long
time  \cite{ZakTa,Ku,MMP},     the  latter
model     is  gauge  equivalent  (in the sense of the completely
integrable systems) to  the NLSE.  Thus,  the NLSE describes the
projection  of the model  on the  tangent space of the spin phase
space. Extending  such a representation to the   2+1 dimensional
systems,  surprisingly,  a  nonvanishing curvature  connection
and the Chern-Simons interaction   appears \cite{CSF}.
\par Let us consider the matrix $\bf {S}$, which   represents  a point
in   the symmetric space
$SU(n+m)/S\lf(U\lf(n\ri) \times U\lf(m\ri)\ri)$   and satisfies   the
constraint
\eq
{\bf {S}}^{2} = {1\over{mn}} I_{n+m} + {(m-n)\over{mn}} {\bf {S}} \virg
\en
where $I_{n+m}$ stands for the identity matrix in
$\lf(n+m\ri)$-dimensions. The  corresponding Heisenberg model is
defined by the  equations of motion
\eq
i\de_{0}\;{\bf {S}} = \frac{mn}{m+n}\;[ {\bf {S}}, \;\nabla^{2}
{\bf {S}} ]                         \label{Hei}
\en
 This matrix can be diagonalized   by a  $U(n+m)$  local
transformation $g$, namely
 \eq
{\bf {S}} = g \Sigma  g^{-1} \virg  \Sigma  = \left(\matrix{{1\over
{n}}I_{n}&0\cr 0& - {1\over {m}}I_{m}\cr}\right) \pu \label{diagS}
\en
 Now, in order  to construct the tangent space representation for
this model we introduce the  chiral current
\eq J_{\mu } = g^{-1} \de_ {\mu } g
= J^{(0)}_{\mu }+ J^{(1)}_{\mu }  \qquad  (\mu  = 0,1,2) \virg
    \label{chircurr}
\en
\noindent where the  ${\bf Z}_{2}$  graduation has been used. In
particular,  $ J_{\mu }$  has the same  block decomposition
presented  in Equation (\ref{aiii}).  Furthermore, by  virtue of
(\ref{chircurr}), the  current $ J_{\mu }$  can be considered as a
  gauge connection   satisfying    the zero-curvature condition,
that is exactly  the classical equation  of motion
(\ref{eqmot})  for the Chern-Simons
topological field theory.  Then, inspired  by  the discussion at
the end of  Section II,
we  can use   the dynamical  Eq.  (\ref{Hei}) in  the
tangent space formulation  as a constraint
for the current components, namely
\eq
i [J_{0},\Sigma ] = \frac{ mn}{m+n}\;( [ \Sigma  ,[ J_{\mu }[ J_{\mu },
\Sigma  ] ] ] + [ \Sigma  , [ \de  _{\mu }  J_{\mu }, \Sigma  ] ] )\pu
\en Thus, the dynamics for the currents  $ J_{\mu }$  is defined by the
zero-curvature equation (\ref{eqmot}). From   the decomposition
(\ref{chircurr})   and  by using the
notation introduced in   (\ref{aiii}), we
get
\eq q_{0} = i(\partial _{j}q_{j} + i v^{(m)}_{j}q_{j} -
iq_{j} v^{(n)}_{j}) \virg
\en
or, equivalently ( see  (\ref{cp}) )
\eq
q_{0} = 2 i \lf(  D\,\psi_{+}  +
\bar{D} \psi _{-} \ri) \label{spinevol}
\en
 where we have introduced the "covariant" derivatives $ D = \de_{
z} + i v^{(m)}\cdot - \cdot i v^{(n)},
\bar{D} = \de_{\bar{z} } + i v^{(m) \dagger}\cdot
- \cdot i v^{(n) \dagger}$.
 Equation (\ref{spinevol}) is an explicit example of the   constraint
(\ref{gf}).  The substitution of  $q_{0}$  in  the action (\ref{3}) gives
the
  formulation of the generalized Heisenberg model (\ref{Hei}) as a
specific symmetric reduction from a pure non-abelian CS model
\bea
&S  = & \frac{k}{  4\pi} {\int_{\Sigma \times {\bf R}}}
\T\, \{  v^{(n)}\de_{0} v^{(n)
\dagger}- v^{(n)\dagger}\de_{0} v^{(n)}+
v^{(m)}\de_{0} v^{(m) \dagger} -
v^{(m)\dagger}\de_{0} v^{(m)} \nn \\&- &
2 v^{(n)}_{0}(\de_{ z } v^{(n) \dagger} -
\de_{\bar{z}} v^{(n)} + i[ v^{(n)},
v^{(n) \dagger}])\nn \\ & -&
 2 v^{(m)}_{0}(\de_{z} v^{(m) \dagger} -
\de_{\bar{z}} v^{(m)} + i[ v^{(m},
v^{(m) \dagger}]) \nn \\    &+   &
\psi ^{\dagger}_{+}D_{0}\psi _{+} - \psi _{+}D^{\dagger}_{0} \psi
^{\dagger}_{+} -
\psi ^{\dagger}_{-}D_{0}\psi _{-} + \psi _{-}D^{\dagger}_{0} \psi
^{\dagger}_{-} \nn \\ &+&
8i ((D\psi _{+})^{\dagger} D\psi _{+} - (\bar{D}\psi _{-})^{\dagger}
\bar{D}\psi _{-} )\}\;dx^{0}  dz  d\bar{z} \label{4}
\eea
with the supplementary condition \eq  \ga  \;
{ \rm \equiv   }
\; D \psi_{+} - \bar{D}
\psi_{-} = 0  \virg \label{geoconstr}    \en  which is  the  Equation
(\ref{eqm4}) and  can be interpreted as the torsion-free condition
for the spin phase space manifold.

 In the  action (\ref{4}) we
have used   $\T$  as  a global symbol  for the trace in  a given
representation  for $U\lf(n\ri)$ and  $U\lf(m\ri)$,  respectively,
and    $D_{0}=
\de_{0} + i v^{(m)}_{0}\cdot - \cdot i v^{(n)}_{0}$ in analogy  with  $
D$ and
$\bar{D}$.

Let us
notice here, that  the tangent space representation  is similar
to the $CP^{1}$
formulation of the $O\lf(3\ri)$ nonlinear $\sigma$ model
\cite{Eich}. But  in this context  the "matter fields"  are identified
with the complex coordinates of the group, while in our
approach $\psi_{\pm}$  are given in terms of the first-order
derivatives of the same quantities.

Now, we know that the classical Heisenberg model
(\ref{Hei})  is  integrable   for  static configurations,  thus we
have relatively few information about the general properties of the
solutions. From this point of view  it is   interesting to look at spin
models which are integrable in (2+1)-dimensions. In  such a case,
the existence of a  Lax pair enables one to linearize the time
evolution of a generic initial configuration of the considered fields
in terms of the corresponding spectral data, via the so-called
Spectral Transform Method \cite{Konob}.  Thus, one may hope to
obtain a great amount of information on the     topological
properties  of the classical solutions. These observations suggest
the existence of  a deep relation between  integrable systems and
TFT.  However, in (2+1)-dimensions the only
known examples of integrable spin models are the
Ishimori system
\cite{Ishi} and the Manakov-Zakharov-Mikhailov-Ward   chiral
field equation
\cite{MZ,Mikh,Ward}. The former system  is related to the
Davey-Stewartson equation
\cite{Lip,kon,noi,mik} and can be deduced as a special case of the
so-called topological magnet model, which has
been studied in  \cite{bilin,topmagP}.
 The   equations of motion for the $SU(2)/U(1)$  topological magnet
 come from the
Lax pair
\bea L_{1}& =& \alpha I \de_{2} +{\bf S} \de_{1}, \nn \\
L_{2}& =& I
\de_{0} + 2 i {\bf S} \de_{1}^{2} + \lf( i \de_{1} {\bf S} - i
\alpha {\bf S}  \de_{2} {\bf S} - \alpha {\bf S} w_{2} + I w_{1}
\ri) \de_{1} \; ,
\eea   where  ${\bf S} = \sum_{ i =1}^{3} S_{i}\,
\sigma_{i} $ (  the   $ \sigma_{i}  $   are  the Pauli
matrices) satisfies the relation  ${\bf S}^{2} = I_{2}$ and, furthermore,
$w_{j}$ is a $U(1)$    connection representing a  velocity field,  with
$\alpha^{2} =
\pm 1$. Commuting
these operators we obtain
\bea i(\de_{0} + w^{j}\de{j}) {\bf {S}} =
\frac{1}{2 } [{ \bf {S}}, \de^{j}\de_{j} {\bf {S}} ]+i
\lf(\de^{1}w_{1} - \de^{2} w_{2} \ri) {\bf S} , \\
\de_{i}w_{j} - \de_{j}w_{i} = -i\; Tr \lf( {\bf {S}} [\de_{i} {\bf
{S}} , \de_{j} {\bf {S}}]\ri),  \label{vorti}
\eea  with the diagonal metric tensor  $g^{ij}=(1, \alpha ^{2})$ on
the flat space $\Sigma $.  Equation (\ref{vorti})  is a
constraint   on the vorticity of $w_{j}$ and it
    is known,  in the theory of the
quantized vortices in the superfluid ${}^{3} He$,
 as the Mermin-Ho relation \cite{Mermin}.
 If the velocity field $w_{j}$ satisfies  the incompressibility
condition
\eq
\de_{1} w_{1} + \alpha^{2} \de_{2} w_{2} = 0,\en
we obtain a ferromagnetic continuum model with non-trivial
background, for  which several vortex  solutions were found and their
dynamics described  in \cite{bilin,dynbil}. Furthermore,
 if  $w_{j}$  can be expressed in terms of   one  scalar
function (the stream function)  $\phi $ as
\eq
w_{1} = \partial _{2}\phi  ,\qquad  w_{2} = \alpha ^{2}\partial _{1}\phi
\label{cond}\en
we obtain
the Ishimori model
\bea
i(\partial _{0}& + &\partial _{1}\phi  \partial _{2} + \partial _{2}\phi
\partial _{1}){\bf {S}} = \frac{1}{2} [{ \bf {S}}, (\partial ^{2}_{1} +
\alpha ^{2}\partial ^{2}_{2}) {\bf {S}} ] \virg
\\ (&\partial ^{2}_{1}& -  \alpha ^{2}\partial ^{2}_{2})\phi
= -i  \alpha ^{2}{ \bf {S}} [\partial _{1}{\bf {S}}
, \partial _{2} {\bf {S}}]\nn
\pu\eea
 In the tangent space representation,  this system   takes the form
\eq
q_{0}= i(D^{j} + i w^{j} )q_{j} \label{ishitan} ,
\en
\eq \lf( \de_{1}^{2} - \alpha^{2}  \de_{2}^{2} \ri) \phi
 = - 4i ( \bar{q}_{1}q_{2} - \bar{q}_{2}q_{1})  \; , \label{auxf}
\en
where $w$ is defined by (\ref{cond}).  Solving  Eq.
(\ref{auxf})  for $\phi$ under suitable boundary conditions,  and replacing
into (\ref{ishitan}),
 we obtain a nonlocal   constraint for $q_{0}$, analogous to
(\ref{spinevol}). In this way we realize a new type of gauge fixing
condition (\ref{gf}), which is integrable.

Finally, in the context of Heisenberg
models  we mention  the possibility to choose a different
constraint of the type (\ref{gf}),  which cannot be solved explicitly  for
$q_{0}$. Indeed, if we   consider  the  generalized  relativistic model
for the  antiferromagnetic  ground state  \cite{Salam}, in the
tangent space representation  it reads
\eq
\partial _{\mu }q_{\mu } +\hbox{ iV}^{(m)}_{\mu }q_{\mu } -
\hbox{ iq}_{\mu }V^{(n)}_{\mu } = 0 \qquad (\mu  = 0,1,2) \pu
\en
 Because of the   complete
space-time  symmetric form of these  equations,  we
can add them as   constraints  to the  action  (\ref{3})  by introducing a
suitable   Lagrange multiplier.  Such  constraints can be interpreted as a
non-abelian Lorentz  gauge condition for  the TFT.

\resection{ Canonical structure   of the
$SU\lf(2\ri)/U\lf(1\ri)$ model}

In this Section, after some  general considerations concerning the
Hamiltonian structure of the theories (\ref{1}) -  (\ref{3}) , we
perform a  more detailed treatment  of  the
$SU\lf(2\ri)/U\lf(1\ri)$ case.

Resorting to   the  "trivialization"  of
the three manifold    ${\cal{M}} \equiv  \Sigma  \times {\bf R} $
( see  Sec. 2)  and following the standard prescriptions
\cite{w2}, we  equip  the general non-abelian CS model
(\ref{1}) with
 the canonical structure  given by
\eq\left\{{
A_{i}^{a}\lf(\bf{x}\ri),{A}_{j}^{b}}\lf(\bf{y}\ri)\right\}={4\pi  \over
k}{\varepsilon }_{ij}{\delta }^{a b}\delta \left({\bf x\rm -\bf
y}\right) \virg
\en where  ${A}_{i}^{a} \; \lf( i= 1, 2\ri)$  are the
spatial components of the connection $J$ (see Eq.
(\ref{comp}))
 represented in a  certain  basis of the Lie algebra  $\hat{g}$.
 Now, it is natural to use the  ${\bf Z_{2}}$-graduation
 (\ref{z2g})  for these canonical variables, expressed  in terms of
the    fields    $ V =  {\it V} \,dz,\; \bar{V} = \bar{\it
V}\,d\bar{z},\;  M = {\it M}  \,dz $ and $\bar {M} = {\it
\bar{M}} \,d\bar{z}$  introduced in (\ref{VMdecomp}). Thus,
the nonvanishing Poisson brackets are
\bea \left\{{{\it
 V}^{a}\lf(\bf{x}\ri),{\bar{ \it V}} ^{b}\lf(\bf{y}\ri)}\right\}
 ={2\pi  i
\over k} {\delta }^{a b}\delta \left({\bf x  - y}\right)\virg  \nn \\
\left\{{{\it M} ^{\alpha}\lf(\bf{x}\ri),{\it \bar{M}}
^{\beta}}\lf(\bf{y}\ri)\right\}={2  \pi  i \over k} {\delta }^{\alpha
\beta}\delta \left({\bf x\rm -\bf y}\right)\virg \label{PB1}\eea
where
$ {\it V}_{i}^{a}, \; {\it {\bar V}}_{i}^{b} $ span   the
subalgebra  $\hl^{(0)}$ . Analogously,   $ {\it M} ^{\alpha},\;
{\it {\bar M}} ^{\beta} $  span the subalgebra    $\hl^{(1)}$.
 This structure supplies a further motivation for the interpretation of
${\it M }$  and ${\it \bar{M}}$ as matter fields interacting with the
  non-abelian CS gauge fields $\it{ V, \;
\bar{V}}$.

The simplest case  pertinent to such theories with abelian gauge
field  is   the
   $SU(2)/U(1)$  model, whose
action is  straigthforwardly derived from (\ref{3}) and
(\ref{aiii}) -   (\ref{cp}):
\bea   S  =   -  \frac{k}{ \pi} \int_{\Sigma  \times {\bf R}}
\left\{{    \half  \epsilon^{\lambda\, \mu\,\nu} \, v_{\lambda}
\de_{\mu}\,v_{\nu} } \right.\hskip 3cm \nn \\
     \left.{ +   i \half  \lf(\psi ^{*}_{+}
{  D}_{0}\psi _{+} - \psi _{+}  \lf({  D}_{0}\psi _{+}\ri)^{*}
  - \psi ^{*}_{-}   {
D}_{0}\psi _{-} + \psi _{-} \lf({
  D}_{0}\psi _{-}\ri)^{*}\ri)  } \right. \hskip
1cm  \label{5}\\
\left.{ - i q^{*}_{0} \lf({   D}\psi _{+} - {\bar{  D}}\psi _{-} \ri)
+  i q_{0} \lf({ D}\psi _{+}  - {\bar{ D}}\psi _{-} \ri)^{*}  }\right\}
 \,dx^{0}\,dx^{1}\,dx^{2}  \nn
\virg
\eea
where $D_{0} = \de_{0} - 2 i v_{0}$, $ D = \de_{z}  - 2 i v $ and
$\bar{ D} = \de_{\bar{z}}  - 2 i v^{*} $  with
 $v = \half \lf( v_{1} - i v_{2}\ri)$.
 From (\ref{PB1}) we obtain the
  corresponding canonical Poisson structure related to (\ref{5}), namely
\bea \left\{\,
v_{i} \lf(\bf{x}\ri),\, {v}_{j} \lf(\bf{y}\ri)\right\} =  - {\pi  \over
k} {\varepsilon }_{ij} \delta \left({\bf {x  - y}} \right) \;,
\left\{{\psi_{\pm} \lf(\bf{x}\ri),\,
\bar{\psi}_{\pm}\lf(\bf{y}\ri)}\right\} = \pm {  \pi  i \over k}
 \delta \left({\bf x  - y}\right)  . \label{PB2}\eea
 In analogy with our observation about the action (\ref{3}),
 in  (\ref{5}) we are considering $v_{0}$ and  $q_{0}$   as
Lagrange
 multipliers, enforcing the constraints of the model, i.e.  the
 GCS
law
 $\Ga_{1}  = \de_{1} v_{2} - \de_{2} v_{1} + 2
\lf({\left|{{\psi }_{+}}\right|}^{2}
 - {\left|{{\psi }_{-}}\right|}^{2}\ri) $,
and the  complex  "torsion-free"  constraint $\ga = {   D}\psi _{+} -
{\bar{  D}}\psi _{-}$, which specializes
 Eq. (\ref{geoconstr})  to the abelian case. They generate the $su\lf(2\ri)$
algebra of the  gauge  symmetry transformations.  Furthermore, the GCS law
constraint $\Ga_{1}$ generates the
$U\lf( 1 \ri)$  subgroup of local gauge transformations. Moreover,
following the Dirac's classification  of the constraints,
 we can
introduce the set of primary constraints   $\Ga_{0} \ \equiv  \pi_{0} = 0$,
$\Ga_{2}  \equiv  \pi_{q} = 0$, where  $\pi_{0}$
is  the  momentum conjugated to the canonical variable
$v_{0}$ and $
\pi_{q}$ is  the  momentum conjugated to $q_{0}$ such that
\eq \{ v_{0}, \pi_{0}\}
=  - \frac{ \pi}{k}
\delta \left({\bf x} - {\bf y} \right)
   \virg \{ q_{0}, \pi_{q}\} =    \frac{  \pi i}{k} \,
\delta \left({\bf x} - {\bf y} \right) \pu \en
Hence,
 we  can write   the
Hamiltonian
$ H = \int_{\Sigma}{\cal H}\;d^{2}x$
  in terms of a set of first-class  constraints	only \cite{Dirac}.
Indeed, we have
\eq
 {\cal H} =   \frac{k}{\pi} \lf(   {v}_{0}{\Ga }_{1}
+ i{q}_{0}{\ga}^{*} - i{q}_{0}^{*}\ga - {f}_{0}{\Ga }_{0} +
 i {g}_{0}{\Ga }_{2} -
i {g}_{0}^{*}{\Ga }_{2}^{*}  \ri)\virg  \label{hamilt}
\en
where  now the
GCS  law	 $\Ga_{1} = \lf\{\pi_{0}, H\ri\}$,
  $\ga =  \left\{{{\pi }_{q}^{*}, H}\right\}$ and
$\ga^{*} =  \left\{{{\pi }_{q}, H}\right\}$
are   to be  considered now as  secondary constraints.
We	see	also that	all constraints
 $\lf\{ \Ga_{1}, \ga \ri\}\otimes
\lf\{\Ga_{0},\Ga_{2}\ri\}$
form	a closed	Lie algebra
		with	abelian	radical,
 and	their	dynamics	is	 also	weakly invariant
( $   \lf\{ \ga, H\ri\}
  \approx	\lf\{ \Ga_{i}, H\ri\} \approx  0$
for $ i=0,1,	2$ ).	Furthermore, one	easily	checks
that the determinant of their
Poisson brackets is  vanishing.
The Hamiltonian itself is weakly vanishing, since it is
a linear combination of constraints.  Finally,  in the expression
(\ref{hamilt})
$f_{0}$,   $g_{0}$ and $g_{0}^{*}$  are
 arbitrary  functions, which  characterize   the evolution of $v_{0}$,
$q_{0}$ and  $q_{0}^{*}$, respectively.

At this stage we recall that,   following the Dirac  ideas
\cite{Dirac}, in
\cite{Boya} it  is suggested to formulate both
classical	and quantum $U\lf( 1
\ri)$ CS theories in 2+1 dimensions
in	a completely invariant	way. 	This result
is achieved 	by introducing in	the
hamiltonian formalism	new	momenta,
which	are	equal to	the constraints	imposed	by
the gauge	theory.	In
this	case,	the corresponding	new	conjugated
coordinates are pure
gauge functions.
	Thus,	one	can provide	a	Hamiltonian
depending	only on
the gauge invariant	degrees	of freedom.
Here	we	show that	this
separation between two	distinguished classes	of
variables cannot	be
performed	up to	the	end,
	at	least	when only	a	gauge	subgroup is
explicitly	solved. To	be	precise,	we
solve the	GCS constraint,
associated with 	the
$U(1)$ symmetry,	and
reformulate	the	Hamiltonian	(\ref{hamilt})	in
terms	of	the
$U(1)$-invariant	degrees	of	freedom	and	of
	pure	gauge	coordinates.
However,	we	cannot	separate
the	Hamiltonian	in	two	pieces,	one	of	which
contains	only	 $U(1)$-invariant
canonical	variables.	This	phenomenon	seems
to	be	connected	with	the	non-abelian
	structure	of	the	algebra	of
constraints	in	a	completely	constrained
	Hamiltonian.	Our	point	of	view	is
that	this	algebra	cannot	be	"strongly
abelianized"	by	using	a	unique
regular	canonical	transformation.

Following the scheme of 	 \cite{Boya},
	first	we restrict ourselves to
the planar geometry taking $\Sigma \equiv {\bf R}^{2}$ and
the rotational	covariance	of	the theory,	by
expressing   $v_{i}$  into
a     longitudinal and  a    transverse part
\eq v_{i}({\bf x}) =   \de_{i}\,\eta ({\bf x}) - \epsilon
_{ij}(\partial ^{-1}_{j}B)({\bf x}) \virg \en  where  we have applied
the	operator	  $\partial^{-1}_{j}f \left({\bf x}\right) =
{1\over{2\pi}}
\partial^{\left(x\right)}_{j}\int
  \ln |{\bf x}-{\bf y}| f({\bf y}) d^{2}y $ to	 the
magnetic field    $ B = \epsilon ^{ij}\partial _{i}v_{j} $ ( notice
that $\de_{1} \de_{1}^{-1} + \de_{2} \de_{2}^{-1} = 1$).
Analogously, it
is convenient to express
$
\psi _{\pm }$ in  terms of the
 canonical variables  $\left(Q_{\pm } ,  P_{\pm }\right)$
by   \eq  \psi _{\pm } = \sqrt{{\pi\over{2 k }} }\left(Q_{\pm }
\mp  i P_{\pm }\right).\en  Then,  the Poisson
structure  is given	by
    \bea  \lf\{Q_{\pm }({\bf
x}),P_{\pm }({\bf y})\ri\}& = & \lf\{ \eta
({\bf x}), \frac{ k}{\pi} B({\bf y}) \ri \} =
\delta ({\bf x} - {\bf y}),  \label{pbq} \\
\lf\{v_{0}({\bf x}), \pi _{0}({\bf y})\ri\} &=&
  \lf\{q_{0}({\bf x}), \pi _{q}({\bf
y})\ri\} =  \delta ({\bf x} - {\bf y}).\label{pbv}\eea  Any other
Poisson  bracket vanishes.  Equations (\ref{pbq}) and (\ref{pbv}) have
been derived  using   the  rescaled   fields
  $- \frac{k}{\pi}\; v_{0}   \rightarrow  v_{0}$,  $\sqrt{\frac{k}{\pi}}\;
q_{0} \rightarrow  q_{0}$ and  $- i \sqrt{\frac{k}{\pi}}\;
\pi_{q} \rightarrow  \pi_{q}$.
\par   Now,  we look  for  a suitable canonical
transformation, such that some of the new momenta are
equal to  the constraints. In such a way, the  coordinates
canonically conjugated to these momenta  have an
arbitrary  time evolution and  are remnants of  the gauge
invariance of the theory.    \par  First,  let  us denote by  $
\lf(\tilde{Q}_{\pm },  \tilde{P }_{\pm }\ri)$,  $\lf(\tilde{v}_{0},
\tilde{\pi
 }_{0}\ri)$, $\lf (\tilde{\eta }, \tilde{\pi}_{1}\ri) $ ,
$\lf(\tilde{q}_{0},
\tilde{\pi}_{2} \ri)$    new  canonically
conjugated coordinates, where  the momenta
$\tilde{\pi}_{i} $  have to be set  equal to
the constraints
$\Ga_{i},\; \; \lf( i = 0, 1, 2 \ri)$.   This can be done
by	means  of  a suitable generating function
\eq
 W = W \left(Q_{\pm },v_{0},\eta ,  q_{0}; \tilde{P }_{\pm
}, \tilde{\pi  }_{0}, \tilde{\pi }_{1},  \tilde{\pi }_{2} \right) \virg
\en  satisfying the set of generalized
Hamilton-Jacobi equations
\eq
\tilde{\pi }_{i}  =
\Gamma_{i}\left( Q_{\pm },v_{0},\eta,   q_{0}; {\partial
W\over{\partial Q_{\pm }}},  {\partial W\over{\partial
v_{0}}}, {\partial W\over{\partial \eta}}, {\partial W\over{\partial
q_{0}}} \right).
\en
 The integrability of this  system is assured by  the
commutation property $\lf\{
\Gamma_{i}, \Gamma_{j}\ri\} = 0$.  In fact,     its general
solution  is given by
\bea
W =  {\tilde{\pi }}_{0} {v}_{0} + {\tilde{\pi }}_{2} q_{0} + {\tilde{\pi
}}_{2}^{*}  q_{0}^{*} +  \frac{k}{\pi} \,\left({\tilde{\pi }}_{1}
+ \frac{k}{\pi} {\tilde{P}}_{-}^{2} - \frac{k}{\pi}\,{\tilde{P}}_{+}^{ 2}
\right) \eta    \nn \\ + \epsilon_{+} \int^{Q_{+}} {\sqrt
{{\tilde{P}}_{+}^{2}-{Q}_{+}^{2}}\,d{Q}_{+}} - \epsilon_{-}
\int^{Q_{-}} {\sqrt {{\tilde{P}}_{-}^{2}-{Q}_{-}^{2}}\,
d{Q}_{-}}  \qquad \lf(
\epsilon_{\pm}^{2} = 1\ri).
\eea
 The
corresponding   canonical transformation  reads
\bea
 \pi _{0}& = &\tilde{\pi}_{0} ,	\; v_{0} =  \tilde{v }_{0}
,	\;\pi_{2} =
 \tilde{\pi}_{2},	\;\pi_{2}^{*} = \tilde{\pi}_{2}^{*},
\;   q _{0} =
\tilde{q}_{0}, \; q _{0}^{*} =
\tilde{q}_{0}^{*},\\
 B &=&
\tilde{\pi }_{1}+ \frac{\pi}{k}\; \lf( \tilde{P}^{2}_{-} -
\tilde{P}^{2}_{+} \ri) ,\qquad
\eta  =  \frac{\pi}{k}\;\tilde{ \eta }, \eea
\bea
Q_{\pm } = \epsilon_{\pm}\,
\left|{    \tilde{ P }_{\pm }     }\right|
\sin \left(\,{\tilde{Q
}_{\pm }
\over{\tilde{P }_{\pm }}} \pm    \frac{2 \pi}{k}
\tilde{\eta }\right) ,\nn \\
 P_{\pm } =
 \epsilon_{\pm}\,
 \left|{    \tilde{ P }_{\pm }     }\right|
 \cos \left(\,{\tilde{Q }_{\pm }\over{\tilde{ P }_{\pm }
}}  \pm    \frac{2 \pi}{k} \tilde{\eta } \right) .\eea
 Now,   we   define
the  new gauge-invariant degrees of freedom
\eq
\Phi _{\pm } = \pm { i \epsilon_{\pm} \over{\sqrt{2}}}
 \left|{    \tilde{ P }_{\pm }     }\right|
\exp
\left(\pm\,  i \,{\tilde{ Q }_{\pm }\over{\tilde{P}_{\pm }}}\right)  =
\sqrt{\frac{k}{\pi}}\; \psi_{\pm} \exp\left( - 2  i \eta\right) , \en
  which fulfill the   canonical brackets
\eq
\{\Phi _{\pm }({\bf x}), {\bar{\Phi }}_{\pm }({\bf y})\}
= \pm \, i \, \delta  ({\bf x} - {\bf y}) .
\en
 In terms of these new variables,  the
Hamiltonian density   becomes
\eq   {\cal H}  =  - v_{0} \pi_{1}+ i q_{0}   {\ga}^{*} - i  q_{0}^{*}
  {\ga}  - f_{0} \pi_{0} - g_{0} {\pi}_{2} -
g_{0}^{*}  {\pi}_{2}^{*} ,\label{hamilt2}\en
where  we have dropped the
 "$\;\tilde{}\;$" for
simplicity.
Furthermore, the constraint $\gamma$ takes the form
\eq \ga \equiv \hbox{\rm  exp}\lf(   \frac{2 \pi i}{k}
 {\eta}Ê\ri) \lf[\lf( \de_{z} - \half \de_{\bar{z}}^{-1}\lf(B\ri) \ri)
\Phi_{+} -
\lf(
\de_{\bar{z}} - \half \de_{z}^{-1}\lf(B\ri) \ri) \Phi_{-}\ri] =
0,
\en
where  the operator  $\de_{\bar{z}}^{-1} \lf(f\lf( z,
\bar{z} \ri)\ri) =
  i \de_{z} \,\int \frac{1}{ \pi} ln \lf| {z - \xi}\ri|  f\lf( \xi,
\bar{\xi} \ri) \, d\xi \w d\bar{\xi}$
   acts on $B = \pi_{1} +
\frac{\pi}{k} \lf(\lf|\Phi_{-}\ri|^{2} - \lf|\Phi_{+}\ri|^{2}\ri)$.
{}From this expression we see that $\ga$ still contains explicitly the gauge
variables $\lf( \eta,\pi_{1}\ri)$ through the exponential
 factor and the expression of $B$ given above.
However,  we cannot extract them from $\ga$  in order to get a completely
$U(1)$-gauge invariant constraint  without  the introduction of second
 class
constraints. Moreover, the equations of motion for $\Phi_{\pm}$, deriving
from the Hamiltonian (\ref{hamilt2}),  still involve $q_{0}$, which
has an arbitrary dynamics. This happens because  $\Phi_{\pm}$ are
$U\lf(1\ri)$-gauge invariant  only.

Moreover, the classical dynamics is restricted
on a manifold of solutions defined by the constraints
\eq \pi_{1} = 0,  \qquad	 \ga = 0 \virg \en
which  become
\bea	B = &
\frac{ 2 \pi}{k} \lf(\lf|\Phi_{-}\ri|^{2} - \lf|\Phi_{+}\ri|^{2}\ri)&
\label{mag}\\
\lf( \de_{z} - 	\half \de_{\bar{z}}^{-1}\lf(B\ri) \ri)
\Phi_{+}& -
\lf(
\de_{\bar{z}} + \half \de_{z}^{-1}\lf(B\ri) \ri)& \Phi_{-}
=  0 \pu\label{cBc}\eea

These equations enable us to look for time independent solutions,
since no  evolution operator is included. On the other hand,  from
the general theory of the completely constrained systems
\cite{gold}, we know that all dynamical informations are encoded into
general canonical transformations, which   can be
considered as gauge transformations and time reparametrizations. In such a
way we can generate time-dependent solutions from the static ones.

 A	special
subcase of
	 Eqs.	(\ref{cBc}), corresponding to the self-dual CS model, is
obtained when
$\Phi_{-}$	 vanishes (or, in	alternative,
$\Phi_{+}$).	In	fact,	by	setting  $\Phi_{+} =
\rho^{\half}\,	e^{i	\chi}$	it easy to see that   the
previous	equations
reduce to the Liouville	equation
\eq
\nabla^{2}\ln\,\rho	=	-	\frac{8	\pi}{k}	\rho,
\en
whose	general
solution
is	given	in	terms	of	an
arbitrary	holomorphic	function
\cite{Liou} $\zeta$ by the relation
\eq
\rho =  \frac{k}{4 \pi} \, \frac{\lf| \de_{z} \zeta \ri|^{2}}{\lf( 1
+
\lf|
\zeta
\ri|^{2}\ri)^{2}} \pu
\en
	The	phase
	$\chi$	is	an	arbitrary
harmonic	function.	 In particular, choosing
$\zeta =
\sum_{n = 1}^{N} \frac{c_{n}}{z-z_{n}}$, where $c_{n}$ and $z_{n}$ are
arbitrary complex numbers, and  assuming that  the phase $\chi$ is  a
regular  and singlevalued  function,  we obtain the so-called self-dual CS
solitons \cite{japi2}. Finally, let us observe that since the
Liouville equation is conformally invariant,  all
reductions we have performed on the original CS model preserve this
special infinite dimensional symmetry group. A	more	general	situation
occurs	when (see
\cite{CSF})
\eq
\lf( \de_{z} - \half  \de_{\bar{z}}^{-1}\lf(B\ri) \ri)
\Phi_{+}	=	\lf(
\de_{\bar{z}} + \half  \de_{ {z}}^{-1}\lf(B\ri) \ri) \Phi_{-}  =  0
\virg	\label{gen-sD}
\en where	both the fields $\Phi_{\pm}$
are	 	different	from	zero. Let us
notice that  the system of these equations with Eq.
(\ref{mag}) corresponds
to the so-called Hitchin equation introduced   as  a  reduction of
the self-dual Yang-Mills equations  in 4 dimensions \cite{Hitchin}.

The	relations	(\ref{gen-sD}) suggest to introduce     a
holomorphic function $U$  (   $\de_{\bar{z}}U	= 0$)  defined by
\eq
	U\lf(	z,	\bar{z}\ri)
=		\bar{\Phi}_{+}\lf(	z,
\bar{z}\ri)\,\Phi_{-}\lf(	z,	\bar{z}\ri). \label{U}
\en	This quantity	is	the analogous	of	the
	holomorphic component of
the energy-momentum	stress	tensor	in the Conformal Field Theories
\cite{BPZ}.	Now,	let	us	suppose
that		U	is	a given entire function.
Then,  	we can solve for instance  (\ref{U}) with respect to
 $\Phi_{-}$
and write down the equations
\bea {\nabla }^{2}\sigma  = - {16\pi  \over k}\;{\rm e}^{\eta
}\,{\rm sinh} \,\sigma \nn \\
{\nabla }^{2} \eta = {\nabla }^{2} \chi =0 \virg \label{affine-sG}
\eea where we have introduced $\sigma = {\rm ln} \,\lf(
\lf|\Phi_{+}\ri|^{2}\,/\, \lf| U \ri|\,\ri)$, $\eta = \hbox{ \rm
Re  ln}\,U$ and $ \chi =
\hbox{arg}\lf(\Phi_{+}\ri)$. Equations (\ref{affine-sG}) are
conformally invariant and  are strictly related to  the integrable
conformal invariant affine Toda field theory  (see
\cite{BabBo}). Actually, this model contains three fields. Two of
them  correspond to $\sigma$ and $\eta$  in Eq.
(\ref{affine-sG}). The third one is completely determined in terms
of the former.
Now, some comments are in order.
Precisely,  in the limit  $\eta \rightarrow 0
$, Eq. (\ref{affine-sG}) becomes the so-called sinh-Gordon equation.
Conversely,  in the limit $\eta
\rightarrow \infty$ we recover  the Liouville equation. All these
models are completely integrable systems and their  solutions can be
studied by resorting to  the Inverse Spectral Transform (IST)
method. In particular, for $k >0$ the sinh-Gordon equation
admits solutions
with pointlike singularities, whose behavior at large distances is
given by $\sigma \rightarrow A \,K_{0}\lf(r\ri)$ , where
$K_{0}$ is the modified Bessel function of  the second kind of order $0$
\cite{Kaup}.However, such a solution cannot be given in closed form
and  is
related to the Painlev\'e transcendents.
Multicharged solutions have been  also
dicussed \cite{Kaup2}.

For  $k <0$  one can obtain  analytic solutions
satisfying the constant boundary conditions on a finite rectangular
boundary. Thus, by using  the IST  method for finite gap solutions,
one can find multiperiodic solutions in terms of the  Riemann
$\theta$ function \cite{Ting}.    On the other hand, concerning the
sinh-Gordon equation, many possible interpretations exist, which are
different from the affine conformal Toda model mentioned above.  For
instance, it can be regarded as  a perturbation of the free massless
conformal model, or as a perturbation of the conformal Liouville
model \cite{BPZ}. It was studied in connection with a
  two-component Coulomb gas \cite{Martinov} or an  anyon-anti-anyon
system
\cite{CSF}. In a different context it was used in the study of  a
system of  vortices  in a bounded magnetized plasma \cite{Book} and
of the counter-rotating vortices in an inviscid and incompressible
fluid
\cite{Mallier}.

Now,  we can figure out that among all possible evolutions
associated with  the model (\ref{5}) we can select some of them
which seem to be particularly interesting from the point of view of
their physical interpretation.  Indeed, instead of using  the Weyl
condition  which in our case  reads
$v_{0} = 0$ and
$q_{0} = 0$, we can  choose the multiplier $q_{0}$  to be precisely  that
prescribed by Eq. (\ref{spinevol}).  This means: i) that the action
will contain  only the fields
$\psi_{\pm}$ and
$v_{\mu}$, ii) in order to recover the topological
symmetry  the constraint $\ga$ must be added  to the  equations
of motion obtained  by  the new action. The system arising in this way
is
 the  $SU\lf( 2\ri) / U\lf( 1 \ri)$ Heisenberg  model in  the tangent
space representation, that is the abelian reduction of model
(\ref{4}). It  contains   the difference of  two Pauli
actions for  non-relativistic charged scalar matter fields
$\psi_{+}$ and $\psi_{-}$,  coupled to  an  abelian  CS
field
$v_{\mu}$.

At this point, first   we notice  that the mapping
between Heisenberg model and CS model,
provided by Eqs.  (\ref{aiii}-\ref{cp})
and (\ref{diagS}-\ref{chircurr}),
  enables us to
express in a natural way   the magnetic field in terms  of the charge
density (see  Section II). Moreover, the magnetic
energy of the Heisenberg system  is   interpreted  directly
 in terms of the
"number of particles"  of   the CS gauged model. In fact,  after
some simple algebraic manipulations, the equations
of motion deriving from expression (\ref{4})
reduce  to a pair of  NLSE  for $\psi_{\pm}$ coupled through the
abelian  CS gauge field and  constrained   by   Eq. (\ref{geoconstr})
(see
\cite{CSF}). This   situation is quite  similar to that of Ref.
\cite{japi1,japi2,japi3}.  In the static case, i.e. in the
"quasi-Weyl" condition $q_{0} = 0$, these  equations reduce to the
self-dual configurations discussed above.
In this context,  classical
 self-dual CS solitons  and  multiperiodic solutions
 can be  interpreted in terms of magnetic bubbles (vortices) in the
spin planar model \cite{self}. These configurations  have spin
angular momentum
 proportional to ${\cal Q}_{el}^{2} $  \cite {pav}.  The vortices
possess topological properties, which are  invariant under  general
gauge transformations  like (\ref{gauge1}).  Then,  we can
speculate  that  vortex solutions  i)  can be found  in the other
spin  models  obtained by a different gauge choice,
  ii)   they    describe a special sector in the moduli space of
the original   TFT  (\ref{1}).

Moreover,   we   found  the Hamiltonian
structure of the  Heisenberg model in the tangent space
representation,   by resorting  to  the symplectic method of
 Hamiltonian reduction \cite{FaJa} addressed to the
gauge-invariant
  approach  \cite{Boya}  which  leads, in this case,
  to a separation of the gauge variables from the gauge-invariant
Hamiltonian  \cite{theor,pav}.  In
terms of the $U\lf( 1\ri)$-gauge invariant degrees of freedom
$\Phi _{\pm }  $,    the Hamiltonian density  is  given
by
\eq
{\cal H}=  4{\left|{\left({{\partial }_{\bar{z}} + \half {\partial
}_{z}^{-1}}\lf( B
\ri)
\right){\Phi }_{-}}\right|}^{2} - 4 {\left|{\left(
{{\partial }_{z} -
\half {\partial }_{\bar{z}}^{-1}
\lf( B \ri)} \right){\Phi }_{+}}\right|}^{2}
-{f}_{0}{\pi }_{0}-{v}_{0}{\pi }_{1},\label{hamiltHei}
\en  where we do not need to introduce
 the conjugated momenta to $q_{0}$,
$q_{0}^{*}$ and the related   primary
constraints. Since the last two terms
in (\ref{hamiltHei}) have vanishing Poisson
brackets with $\Phi_{\pm}$, they
can be  set strongly equal to zero,
when we study the time evolution of the
gauge-invariant degrees of freedom given by
${\dot{\Phi }}_{\pm }=\left\{{{\Phi }_{\pm },H}\right\}$.

By this procedure we have selected a particular time evolution, among all
reparametrization transforms admitted by the pure $SU \lf( 2 \ri)$ model
(\ref{5}). But we could repeat the same argument using a different
choice of
$q_{0}$, for instance the expression (\ref{ishitan}) for the Ishimori
model.
The integrability of this model will provide a complete description
of the
corresponding phase space. Thus we expect that this  could improve
the study
of the original TFT.

\resection{ Quantization of the $SU\lf(2\ri)/U\lf(1\ri)$ model}

 The quantum theory of the model (\ref{5}) can be carried out   by
means of  the correspondence
${\Phi}_{\pm }\rightarrow {\hat{\Phi}}_{\pm }$ and
${\bar{\Phi}}_{\pm }\rightarrow {\hat{\Phi}}^{\dagger}_{\pm }$  and
replacing the canonical brackets  by  the equal-time commutators in
the boson case (or the anticommutators in the fermionic case)
 \eq [{\hat \Phi} _{\pm}({\bf x}),{\hat{\Phi }}^{\dagger}_{\pm}({\bf
y})] = \mp\,
\delta ({\bf x} - {\bf y}) .\label{comm}\en
The presence of a different signature in  Eq. (\ref{comm})
leads   generally  to an   unbounded quantum energy spectrum,
whose    treatment  requires some special care. However, we know
that the classical value of the energy is zero  because of the
completely constrained character  of the classical Hamiltonian.
So we expect that all the  physical quantum states have to
be eigenstates  corresponding to  the energy eigenvalue $0$.
 Furthermore, the
first class constraints
$\Ga_{i}$ become  the operators $\hat{\Ga}_{i}$, which must
annihilate the physical states. In particular,  the
operator  $\hat{\Ga}_{1}$ associated with  the GCS law  has to
annihilate  the physical states. This  implies that such states are
independent  from
$v_{0}$ and are invariant under time-independent gauge
transformations. Therefore, all the operators   $\hat{\Ga}_{i}$
must commute among themselves. Finally,  we have to associate
with   the constraint
$\ga$ a corresponding operator  $\hat{\ga}$, which also
annihilates  the physical states.

 For the specific case  of  the Heisenberg model, in the
subspace  of  the physical states,     we can write down a quantum
Hamiltonian involving only the operators $\hat{\Phi}_{\pm}$ and
their hermitians:
   \eq
\hat{H}=4\int_{}^{}\left\{{{\hat{\Phi }}_{+}^{\dagger}{\left({{\partial
}_{z} - \half {\partial
}_{\bar{z}}^{-1}\left({\hat{B}}\right)}\right)}^{2}{\hat{\Phi
}}_{+}-{\hat{\Phi
}}_{-}^{\dagger}{\left({{\partial }_{\bar{z}} - \half {\partial
}_{z}^{-1}\left({\hat{B}}\right)}\right)}^{2}{\hat{\Phi }}_{-}}\right\}dz
d\bar{z},
\label{qham}\en
 where $ \hat{B} = \frac{\pi}{k} \lf( {\hat{\Phi }}_{-}^{\dagger}
{\hat{\Phi }}_{-}  - {\hat{\Phi }}_{+}^{\dagger} {\hat{\Phi }}_{+}
\ri)$,  and  the normal ordering of the operators is used. The
quantized free-torsion constraint
$\hat{\ga}$ takes the form
\eq
\hat{\ga} = \lf( \de_{z} -
 \half  \de_{\bar{z}}^{-1}\lf( \hat{B}\ri)\ri)\hat{\Phi}_{+} -
\lf( \de_{\bar{z}} - \half
 \de_{z}^{-1}\lf( \hat{B}\ri)\ri)\hat{\Phi}_{-} \pu
\en
  Therefore, since  $\lf[ \hat{B}\lf({\bf x }\ri), \hat{\eta}\lf({\bf y
}\ri) \ri] = - \frac{\pi i}{k} \delta ({\bf x} - {\bf y})$ holds,
 one has
the relation
\eq \exp \left(i\,{\hat \eta}  \left({\bf
y}\right)\right){\hat B}\left({\bf x}\right)\exp
\left(-i\,{\hat \eta}  \left({\bf y}\right)\right) = {\hat
B}\left({\bf x}\right) -  \frac{\pi }{k}  \,\delta ^{2}\left({\bf x}
- {\bf y}\right).
\en
This result is exploited
to prove    that $ {\hat \Phi}_{\pm} $  are  the
$U\lf( 1 \ri)$-gauge  invariant   operators,   which    create  a
charge-solenoid composite,  having magnetic  flux  equal to  $\mp
\pi / k $.

 Now, we define the  quantum vacuum state    by  the relation
${\hat
\Phi} _{\pm }|{\bf 0}> = 0$. Thus we can introduce  two different
  particles  number operators
 \eq {\hat N}_{\pm }=
\int \, {\hat
\Phi} ^{\dagger}_{\pm }{\hat \Phi} _{\pm }\,d ^{2} x \pu\en
  These operators commute  between themselves and  with the
Hamiltonian operator (\ref{qham}). Thus, we can   formally
construct the  common eigenstates of the energy and  of the
occupation numbers for both types of  particles by
\bea  |N_{+},
N_{-}>&  = \int_{}^{}\prod\nolimits\limits_{i=1}^{{N}_{+}}
{d}^{2}{x}_{i}^{+}\prod\nolimits\limits_{j=1}^{{N}_{-}}
{d}^{2}{x}_{j}^{-}\Psi \left({\bf
x}^{+}_{1},\dots  , {\bf x}^{+}_{N_{+}},{\bf x}^{-}_{1},\dots  ,
{\bf x}^{-}_{N_{-}}\right) \nn \\  &{\hat \Phi}
^{\dagger}_{+}\left({\bf x}^{+}_{1}\right)\ldots  {\hat \Phi}
^{\dagger}_{+}\left({\bf x}^{+}_{N_{+}}\right) {\hat \Phi}
^{\dagger}_{-}\left({\bf x}^{-}_{1}\right)\ldots  {\hat \Phi}
^{\dagger}_{-}\left({\bf x}^{-}_{N_{-}}\right) |{\bf 0}>.
\label{state}
\eea
 The function  $\Psi $ is an
arbitrary element of  the Hilbert space  ${\sl L}_{2} [{\cal
R}^{2\left(N_{+} + N_{-}\right)}]$  obeying   the
Schr\"odinger equation for   $\left(N_{+}
+N_{-}\right)$-bodies.

Now, to be physical,  the state (\ref{state}) has to be also
annihilated by $\hat{\ga}$.  For    states  with  finite number of
particle we can find  exactly solvable equations for $\Psi
$    only when one type of particles is present. For instance,
the
$\left|{{N}_{+},0}\right\rangle$  state is described  by the bosonic wave
function
\eq {\Psi }\left({{\bf x}_{\rm 1}^{+},\cdots,{\bf x}_{\rm
{N_{+}}}^{+}}\right) = {\cal F}
\left({{\overline{z}}_{1}^{\;+},\cdots,{\overline{z}}_
{{N}_{+}}^{\;+}}\right) \prod\nolimits\limits_{i\,<\,j}
{\left|{{\bf x}_{\rm i}^{+}-{\bf x}_{\rm
j}^{+}}\right|}^{- \,\frac{2 }{k} }\virg \label{wf}
\en   where $\cal F$ is an arbitrary
holomorphic function of its arguments. By
using a singular gauge transformation
\cite{Wil2,Gir},  the expression (\ref{wf})   takes the  form
 of the Laughlin multivalued anyonic  wave function  \cite{Lau2}
\eq {\Psi }\left({{\bf x}_{\rm 1}^{+},\cdots,{\bf x}_{\rm
{N_{+}}}^{+}}\right)= \tilde{\cal F}
\left({{\overline{z}}_{1}^{\;+},\cdots,{\overline{z}}_
{{N}_{+}}^{\;+}}\right)\prod\nolimits\limits_{i\,<\,j} {\left({{
z}_{\rm i}^{+}-{z}_{\rm j}^{+}}\right)}^{- \,\frac{ 1}{k} }\pu \label{wf1}
\en
This wave function acquires   the anyonic phase
${\rm exp } \lf( \frac{i \pi}{k}\ri)$ after the exchange of two
particles.

Then, if we consider the model (\ref{qham}) as  a quantum version of
the magnetic bubble system, we see that it  behaves as  a quantum
anyon system.
Furthermore,  the  wave  function  (\ref{wf1}) can be employed to
describe  a condensate state of bosonic solitons and  may be
related to  the  quantum disordered state of the original
ferromagnet
\cite{zee}. Actually, the previous wave function is not
normalizable on the plane. Its normalization can be obtained by
 introducing an external magnetic field in our
topological model. Normalizable eigenstates can be obtained also
on a compact surface
$\Sigma$ without introducing  an external field \cite{Iengo}. However,
in such a case the anyon gauge does not seem very fruitful.

\resection{Conclusions}

We have studied the classical non-abelian  CS
model, where the gauge fields satisfy certain  geometrical
requirements. In particular, if the gauge fields belong to a
$Z_{2}$-graded Lie algebra,    we are led in a natural way to a
decomposition of the action, in which  a set of matter
fields interacts via  a  (generally non-abelian)  CS field.
The models, described by the action with this property,
  are compared with the
systems arising from the
 so-called tangent space representation of the generalized
Heisenberg models in 2+1 dimensions. From this analysis we
 see that some  special
planar spin models can
be obtained by  a  generally  non-abelian CS theory via a
convenient  gauge-fixing condition. For example, both  the
well-known $SU\lf( 2\ri)$ Heisenberg model and the Ishimori model,
which is completely integrable, can be associated with the TFT  in
special gauges.   The symplectic structure of these models has
been  investigated  mainly  within the gauge-invariant  procedure.
Thus,  we can provide also the quantum theory of the models. In
particular, the many-body wave function is multivalued, since in
the anyonic gauge  it gets  the  phase ${\rm exp } \lf( \frac{i
\pi}{k}\ri)$ by exchanging two particles. The corresponding anyons
have spin  $ s = \frac{1}{2 k} \lf( {\rm mod}\; {\bf Z}\ri)$.

Now, our approach to  TFT  needs further developments
and  several interesting applications are in  order.

First of all, the possibility to handle   nonlinear completely integrable
systems in the context of the quantum TFT  is not completely exploited. A
suitable BRST approach has to be applied in order  to restore the general
covariance. However,  our concern is that of
finding  new results, or also old results in a more easy way, just
working with integrable structures.

 Since the  models discussed in this paper come from a quite
abstract framework,  it is interesting to see whether some of them  play
any  special  role  in  the
  domains of  planar  physics,
where   non-local interactions  of the   CS type   become
important. At present, we  notice  only that in our    models the
coupling constants of the theory  are fixed by the geometry of the
original non-abelian pure CS model and by the constant $k$, which  is an
integer from the quantum theory.  Thus, our models  are special cases
of  many other CS-gauge field theories (see
\cite{bak}-
\cite{japi3}), in which   the  above mentioned
coupling constants can take arbitrary real values and,
 for this reason, can
be considered as pertubations or deformations of ours.

Other  possible physical applications of these ideas concern the
description of the multi-layer Hall systems, treated as planar
pseudospin ferromagnets \cite{doubLay}
  and the study of the quantum Hall fluids in terms of
boundary excitations and of CS topological field theories \cite{Fro}.

 Furthermore, since  the CS TFT is exactly solvable in the
$SU \lf( 2
\ri)$ case  for any
 three-manifolds \cite{w2}, we  speculate that our procedure
should lead to a field theoretical description of an anyonic system
on an arbitrary Riemann surface $\Sigma$ (sphere, torus and higher
genus surfaces).  In these cases the  operator $\de_{j}^{-1}$
  must  be modified,
in order to include a residual topological interaction \cite{Randj}.
However, when these systems  are obtained  by the described procedure,
their solvability is assured by  the above mentioned general
property.

It is also physically interesting to  consider
 the extension of our procedure
  to    noncompact  groups. In fact, for  the $ISO(2,1)$ group the
corresponding  model  is equivalent to   the TFT studied  by Witten  in
connection with the (2+1)-dimensional quantum gravity
\cite{wittgr}.

 Furthermore, our procedure can be used for classifying classical
topological field models by the point of view of their integrability
property.
 This  corresponds to classify the
supplementary constraints,  by requiring that the system of equations
 (\ref{eqm1}) -
(\ref{eqm6}) and  (\ref{gf}) allows  a Lax pair, in analogy with
the procedure and the results contained in
\cite{AthFo}. It is remarkable that in our scheme,  either integrable and
non-integrable gauge-fixing conditions are found. The question  of the
existence of a    relationship between  these two type of structures in the
unifying framework supplied by the original non-abelian CS model is still
open.
 A hint for solving this problem may be found  in the gauge
transformations generated  by $\lambda ^{(1)}$, which   mixes
 the components of the chiral current by the relations
 \bea
 \delta J^{(0)}  = & [J^{(1)},\lambda ^{(1)}]
 \nn \\
 \delta J^{(1)}  = &[J^{(0)},\lambda ^{(1)}] + d\lambda
^{(1)} \pu      \label{gaugeonmatt}
 \eea
 We notice  that  the transformations (\ref{gaugeonmatt}) imply  a non
trivial mixing among old matter and gauge fields.

 Finally,  we point out that in a similar fashion one can build up more
general models,    by considering  non-symmetric spaces.  For
instance,  models involving
$N$ abelian CS fields   can be constructed in a reductive
homogeneous space \G /\cH , where \cH $= U\lf(
1\ri)^{N}$  \cite{Dub,Hel}. Another possible generalization can be obtained
just  considering an higher dimensional $SU \lf( n \ri)$ group. In this
 case we
can  solve
 the GCS laws related to the maximal abelian subgroup of local $U \lf(
1 \ri)$  gauge symmetries.  These theory   should  be solved
 in terms of the Toda models, at least for the
self-dual reduction \cite{jakpiSupp}.
\vskip 1cm

{\large {\bf Aknowledgements}}

 The authors are grateful  to  L.
Bonora, V. Ja. Fainberg, J. Fr\"ohlich, S. Randjbar-Daemi, I.
Todorov, I. V. Tyutin and G. Vitiello for  very helpful discussions.

This work was supported in part by MURST
of Italy and by INFN - Sezione di Lecce.
 One of the authors (O. K. P.) thanks
the Department of Physics of Lecce University for the warm hospitality.
 He  thanks also  TUBITAK
of Turkey and  M. Idemen for supports and hospitality
at the Marmara Research Center.

\vfill
%%%%%%%%%%%%%%%%%%%%%%%%%%%%%%%%%%%%%%%%%%%%%%%%%%%%%%%%%%
\newpage
%%%%%%%%%%%%%%%%%%%%%%%%%%%%%%%%%%%%%%%%%%%%%%%%%%%%%%%%%%%

\end{document}